\documentclass[fleqn,usenatbib]{mnras}

\usepackage{newtxtext,newtxmath}

\usepackage[T1]{fontenc}

\DeclareRobustCommand{\VAN}[3]{#2}
\let\VANthebibliography\thebibliography
\def\thebibliography{\DeclareRobustCommand{\VAN}[3]{##3}\VANthebibliography}

\usepackage{graphicx}	
\usepackage{amsmath}	
\usepackage{siunitx}    
\usepackage{hyperref}   
\usepackage{multirow, makecell}   
\usepackage[table]{xcolor}     
\usepackage{color, soul}
\usepackage{overpic}

\newcommand{\lcdm}{$\Lambda$CDM}
\newcommand{\dmb}{\ensuremath{\Delta m_\mathrm B}}
\newcommand{\pc}{\textsc{PolyChord}}
\newcommand{\blackjax}{\textsc{BlackJAX}}
\newcommand{\hordoverlay}[3][0.95]{
    \begin{overpic}[width=#1\columnwidth]{#3}
        \put(0,0){\includegraphics[width=#1\columnwidth]{#2}}
    \end{overpic}
}
\DeclareSIUnit\parsec{pc}

\title[Dynamic or Systematic?]{Dynamic or Systematic? Bayesian model selection between dark energy and supernova biases}

\author[A.N.~Ormondroyd et al.]{
    A.N.~Ormondroyd,$^{1,2}$\thanks{E-mail: ano23@cam.ac.uk}
    W.J.~Handley,$^{2,3}$
    M.P.~Hobson,$^{1}$
    A.N.~Lasenby,$^{1,2}$
    and D.~Yallup$^{2,3}$
    \\
    $^{1}$Astrophysics Group, Cavendish Laboratory, J.J.~Thomson Avenue, Cambridge, CB3 0HE, UK\\
    $^{2}$Kavli Institute for Cosmology, Madingley Road, Cambridge, CB3 0HA, UK\\
    $^{3}$Institute of Astronomy, Madingley Road, Cambridge, CB3 0HA, UK\\
}

\date{Accepted XXX. Received YYY; in original form ZZZ}

\pubyear{2025}

\begin{document}
\label{firstpage}
\pagerange{\pageref{firstpage}--\pageref{lastpage}}
\maketitle

\begin{abstract}
    The Dark Energy Survey 5-year (DES-5Y) supernovae, combined with Dark Energy Spectroscopic Instrument (DESI) baryon acoustic oscillations, appear to favour Chevallier--Polarski--Linder $(w_0, w_a)$ dynamical dark energy over \lcdm{}.
    It has been suggested in other work that this is driven by a systematic in the DES pipeline, which particularly affects the low-redshift supernovae brought in from legacy surveys.
    It is difficult to investigate these data in isolation, however, as the complicated supernovae pipelines must properly account for selection effects.
    In this work, we discover that a magnitude offset between the low- and high-redshift supernovae is favoured by the Bayesian evidence over the flexknot dark energy found in our previous work.
    In addition, we find that the possible tension between DES-5Y and DESI is significantly reduced by such an offset.
    The recent DES-Dovekie recalibration partially alleviates this tension, but does not eliminate it.
    We also take the opportunity to trial Nested Bridge Sampling with Sequential Monte Carlo as an alternative method for calculating Bayes factors.
\end{abstract}

\begin{keywords}
    methods: statistical -- cosmology: dark energy, cosmological parameters
\end{keywords}



\section{Introduction}

Despite the successes of the standard cosmological model, known as \lcdm{}, the nature of dark energy has remained an enigma for almost three decades \citep{SupernovaSearchTeam:1998, SupernovaCosmologyProject:1998}.
This has motivated the exploration of alternative phenomenological hypotheses, ranging from a first-order expansion, Gaussian processes, and our previous flexknot reconstruction. Ultimately, these approaches all seek evidence that the dark energy equation of state parameter, $w$, has not been $-1$ for all cosmic time \citep{einstein1917, einstein1917centenary}.
Flexknots represent $w(a)$ as a linear spline between a variable number of free-moving knots, encompassing $w$CDM and CPL as special cases with one and two knots respectively \citep{paper1}.
The $n$ knots are indexed from $0$ at the present day, so that $w_0$ is consistent with the usual CPL labelling, and $w_{n-1}$ is the value of $w(a)$ at the earliest time.

\citet{georgedes5y} claims that there is a systematic offset between the distance moduli of low- and high-redshift data in the Dark Energy Survey 5-year (DES-5Y) Type Ia supernovae \citep{des5y}.
DES \citep{vincenzi} have responded to this, reporting that this claim is unsubstantiated and does not properly account for the complicated nature of Type Ia supernovae standardisation.
\citet{baovssnevidence} investigated discarding the mutual supernovae between DES-5Y and Pantheon+ \citep{pantheonplus}, then performed an offset between the low- and high-redshift supernovae in DES-5Y, and found that a cosmological constant is less strongly excluded.
In this work, we ask: is there any \textit{Bayesian} evidence for this?

Since the original submission of this work, DES have released the Dovekie update \citep{dovekie}, which re-calibrates the DES-5Y sample with improved photometric cross-calibration, a retrained SALT3 light curve model, and corrections to the host galaxy colour law. We revisit our analysis accordingly.

We extend the flexknot dark energy reconstructions from \cite{paper1, paper2} with an additional low-redshift magnitude offset parameter, and investigate how this additional degree of freedom affects the Bayesian evidence for \lcdm{}, the CPL \citep{Chevallier:2000qy, Linder:2002et} parameterisation, and our free-form flexknot reconstructions.
We will also examine how the tension ratio is affected by this offset.

This paper is organised as follows.
Section~\ref{sec:context} will outline the wider context of this work.
In Section~\ref{sec:data}, the additional offset parameter in the DES-5Y supernovae likelihoods will be explained.
Section~\ref{sec:methods} will offer a brief recap of the reconstruction method employed in this work, and demonstrate nested bridge sampling as an alternative method for producing posterior samples and Bayes factors.
Results using DES-5Y will be discussed in Section~\ref{sec:results}, updates in light of DES-Dovekie in Section~\ref{sec:dovekie}, and our conclusions in Section~\ref{sec:conclusions}.

\section{Context}\label{sec:context}

Type Ia supernovae are a crucial tool to search for evidence of deviation from \lcdm{}, and are known as ``standard candles'' --- in fact, \textit{standardisable} candles would be a more appropriate moniker, as meticulous calibration work has to be done by collaborations such as DES to create a standardised dataset suitable for use by others to constrain their favourite alternative cosmologies.

\citet{georgedes5y} suggests that the detection of evolving dark energy by the combination of the second release DESI baryon acoustic oscillation (BAO) and DES-5Y Type Ia supernovae is driven by a systematic in the DES pipeline.
In particular, it was suggested that the low-redshift supernovae included in the DES pipeline which are \textit{not} from DES's own measurements, but from the CfA \citep{Hicken2009, Hicken2012}, the Carnegie Supernova Project \citep{Krisciunas2017, Krisciunas2020} and the Foundation Sample \citep{foundation}, have an apparent magnitude $m_\mathrm B$ which is systematically 0.04 magnitudes above both the same supernovae (SNe) in Pantheon+, another supernova dataset, which is apparent when both are compared to the Planck best-fit cosmology.
\citet{vincenzi} responded to this claim, citing analysis improvements compared to Pantheon+, and explained how different selection criteria between the two catalogues means it is expected that the data sets contain differences.
DES have since released the Dovekie re-calibration \citep{dovekie}, which addresses the low-redshift calibration with updated photometry and a retrained SALT3 model, and weakens the preference for evolving dark energy.

The DES-5Y supernova sample represents the largest single-survey dataset of its kind, with over 1600 photometrically classified SNe Ia from the DES programme.
This sample also includes 194 SNe at low redshifts from a number of external surveys to serve as a cosmological anchor.
Broadly, the external supernovae are at redshifts of less than $0.1$, and the DES SNe are from $0.1$ to $1.13$.
In contrast to Pantheon+, DES-5Y presented a stronger preference for evolving dark energy.

Supernova surveys are magnitude-limited, that is, bluer and longer events are more likely to be detected than redder, shorter events: the Malmquist bias.
This bias must be corrected for on a supernova-by-supernova basis, which is done using simulations.
Of course, there are differences between the Pantheon+ and DES-5Y pipelines, which are covered in detail in the appendix of \citet{vincenzi}.
For example, it was found that replacing the \textsc{SALT3} light curve fitting model \citep{Kenworthy2021, taylor2023} of DES-5Y with the older \textsc{SALT2} \citep{salt2} model used by Pantheon+ would have halved the offset found by \citet{georgedes5y}.
DES-5Y and Pantheon+ also use different selection functions, which means that the bias corrections \textit{should} be different, and equivalence of the ``same'' supernova event should not be expected. In fact, Pantheon+ contains a strongly biased selection of the DES-3Y supernovae: those with spectroscopic follow-up.
\citet{baovssnevidence} compare DES-5Y and Pantheon+, with the mutual supernovae excluded, of course, but this does not properly account for selection effects. If it were to be done properly, the mutual supernovae should be deleted before bias corrections.
The approach here suffers a similar limitation, but in lieu of a viable alternative, we proceed.

Two things may be true at the same time. There can be a systematic issue with the low-redshift external supernovae in DES-5Y, even if expecting them to be identical to Pantheon+ is an oversimplification.
Therefore, we seek to investigate whether there is any Bayesian evidence for such an offset \citep{bayes1763essay}.
Deliberately, we do not include the Pantheon+ supernovae in this work.
That way, any suggestion we find for this offset is entirely independent of comparisons between the two pipelines.\footnote{Of course, if it were not for the suggestion of an offset between the two, this work would not have been carried out. It is left as an exercise to the reader to choose an appropriate prior given they are reading this paper.}

\section{Data}\label{sec:data}

In this work, an agnostic approach is taken.
Rather than relying on the -0.04 value as chosen in \citet{georgedes5y}, we add an additional parameter to our likelihood, \dmb{}, which is an offset applied only to the non-DES supernovae:
\begin{equation}
    \begin{aligned}
        \mathcal L(D | \theta) &= \frac{1}{\sqrt{|2\pi\Sigma|}}\exp - \frac 1 2 \mathbf{\Delta}^T\Sigma^{-1}\mathbf{\Delta}\text,\\
        \mathbf{\Delta} &= (\mathbfit m_\mathrm B + \mathbfit s\dmb{} - M_\mathrm B) - \mu(\mathbfit z, \theta)\text.\\
    \end{aligned}
\end{equation}
$\mathbfit s$ is a binary selection mask which is 1 if the corresponding supernova was not from the DES catalogue itself, and 0 if it was from DES.
Setting $\dmb=0$ is equivalent to the standard supernova likelihood.
The distance modulus $\mu$ is calculated from the luminosity distance:
\begin{equation}
    D_\mathrm L(z) = (1+z_\mathrm{hel})c\int_0^{z_\mathrm{HD}}\frac{\mathrm dz'}{H(z')}\text{,} \quad \mu(z) = 5\log_{10}\left(\frac{D_\mathrm L(z)}{\SI{10}{pc}}\right) \text.
\end{equation}
If \dmb{} is unsubstantiated, then it will be Occam-penalised; if the Bayesian evidence is significantly greater with this parameter, however, this is consistent with systematic offset between the supernovae from DES.

Since the absolute magnitude $M_\mathrm B$ is also a parameter to be fitted, an overall offset in $m_\mathrm B$ has no effect on cosmology, so one would obtain the same results with the opposite mask, albeit \dmb{} would have the opposite sign.
In fact, it is possible to marginalise out \dmb{} analytically, this is discussed further in Appendix~\ref{apx:marginalisation}, but all of the results simply sample the parameter.

For a more complete reconstruction of the expansion history, likelihoods with and without the \dmb{} offset were combined with the second major data release of DESI DR2 BAO measurements.
Since release two is the current state of the art, we will refer to this simply as ``DESI BAO'', or just ``DESI'', for brevity.

\section{Methods}\label{sec:methods}

\subsection{Flexknot dark energy reconstructions}
The ``flexknot'' approach of reconstructing the dark energy equation of state parameter is explained in detail in \cite{paper1} and \cite{paper2}, but we will outline the approach again here.
Flexknots are a free-form, model-independent method for reconstructing one-dimensional functions, in this case, $w(a)$.
They consist of a linear spline between $n$ nodes, or ``knots'', whose positions are parameters of the model to be fitted using nested sampling.
The horizontal coordinate of the left- and rightmost knots are fixed, in this case, at zero and one respectively, and the remaining coordinates are free to vary within these bounds, with the restriction that they remain sorted.
The number of knots $n$ is also a parameter of the model; in practice, separate nested sampling runs are performed for each $n$ using \textsc{PolyChord} \citep{polychord1, polychord2}, and the posterior of $n$ is proportional to the evidence for each run.
This is a well-established technique across many areas of cosmology beyond the dark energy equation of state \citep{paper1, paper2, sonke, devazquez}, including the primordial power spectrum \citep{pkhandley, pkvazquez, pkknottedsky, pkcore, pkplanck13, pkplanck15}, the cosmic reionisation history \citep{flexknotreionization, heimersheimfrb}, galaxy cluster profiles \citep{flexknotclusters}, and the $\SI{21}{\centi\metre}$ signal \citep{heimersheim21cm, shen}.
A similar approach was one of the methods used by \cite{2025RPPh...88i8401M}, though with a polynomial of degree $n-1$ through all the knots, rather than separate linear segments.
\begin{table}
    \centering
    \rowcolors{2}{}{gray!25}
    \begin{tabular}{|l|c|}
        \hline
        Parameter & Prior \\
        \hline
        \dmb{} & $[-0.1, 0.1]$ \\
        $n$ & $[1, 20]$ \\
        $a_{n-1}$ & $0$ \\
        $a_{n-2}, \dots, a_1$ & sorted($[a_{n-1}, a_0])$ \\
        $a_0$ & $1$ \\
        $w_{n-1}, \dots, w_0$ & $[-3, 1]$ \\
        $w_a$ & $[-3, 2]$, $w_0+w_a<0$ \\
        $\Omega_\mathrm m$ & $[0.01, 0.99]$ \\
        $H_0r_\mathrm d$ (DESI)& $[3650, 18250]\unit{\kilo\metre\per\second}$ \\
        $H_0$ (Ia) & $[20, 100]\unit{\km \per \s \per \mega \parsec}$ \\
        \hline
    \end{tabular}
    \caption{
        Cosmological priors used in this work.
        Knots are indexed from $w_0$ (today) to $w_{n-1}$ (earliest time).
        Fixed values are indicated by a single number, while uniform priors are denoted by brackets.
        As BAO depend only on the product $H_0r_\mathrm d$, and supernovae depend on $H_0$, the former is sampled only when DESI is included, and the latter is analytically marginalised out.
        Similarly, $w_a$ is used only for the CPL model, with the restriction that the value of $w$ is negative at early times, $w_0 + w_a<0$, as used by other work.
    }
    \label{tab:priors}
\end{table}

Flexknots also have the advantage that $n=1$ and $n=2$ cases correspond to the $w$CDM and CPL models respectively, though, in the latter case, separate sampling runs are also performed with priors consistent with other works for completeness.
In response to comments from presenting our previous work, one small change has been made which was previously shown in an appendix of \citet{paper2}; that is, the upper limit of the prior of the $w$-coordinates of the knots is now $1$. This is more consistent with typical priors used for CPL (e.g: \citet{planck15parameters, planck18vi, desivi, desi2i, desi2ii, desi2de}), and means that \lcdm{} is in the centre of the prior, though of course this is of no consequence for a uniform prior.
Under each reconstruction, the Kullback--Leibler divergence (KL divergence, \citet{dkl}) from the prior to the posterior of $w$ is shown as a function of scale factor or redshift as appropriate.
This quantifies how much the data have compressed the functional prior at each point in cosmic history, and is a more robust measure of constraining power than simply comparing contours.

\begin{figure}
    \begin{center}
        \includegraphics[width=0.48\textwidth]{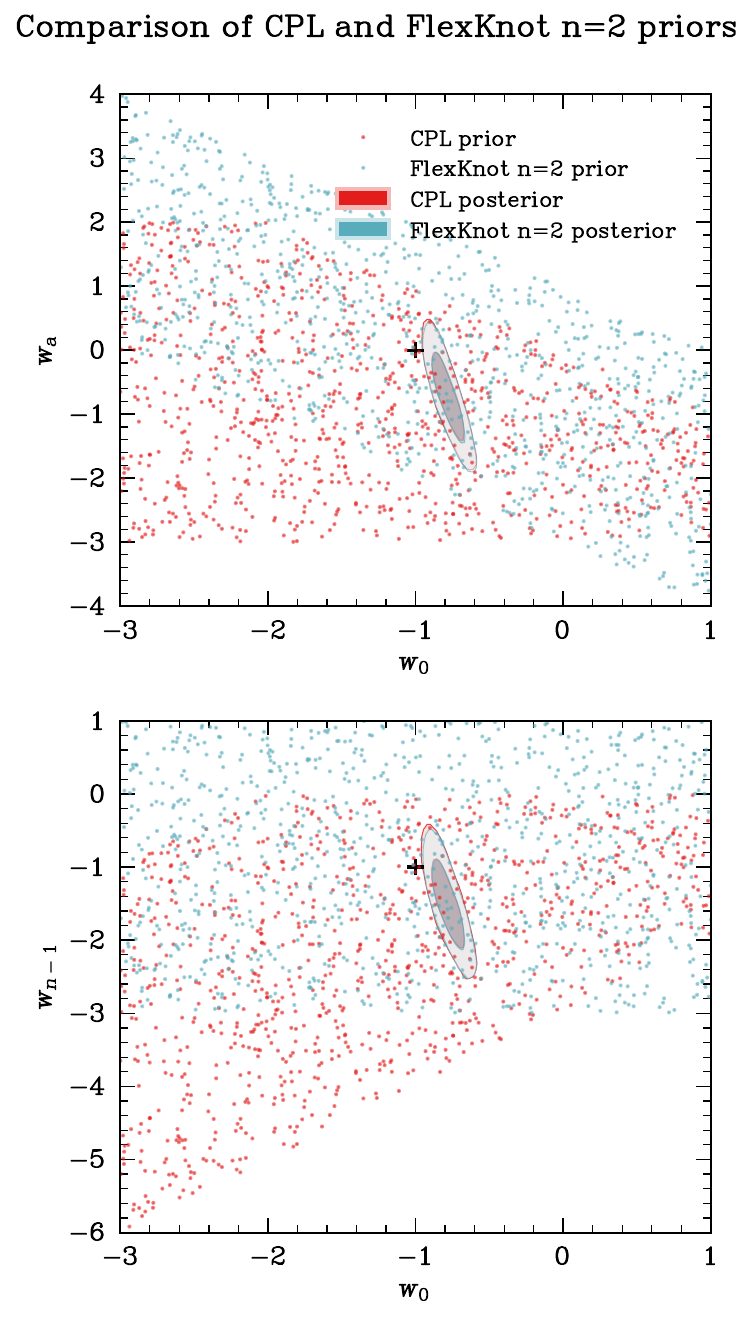}
    \end{center}
    \caption{
        Prior and posterior of CPL and $n=2$ flexknot, using DES-5Y combined with DESI BAO.
        The top panel shows the $(w_0, w_a)$ projection, the lower panel shows $(w_0, w_{n-1})$.
        Prior samples are shown as a scatter, the posteriors are shown as kernel density estimates.
        For reference, the cross marks \lcdm{}.
        The two posteriors are so similar that it is challenging to see one on top of the other!
    }\label{fig:prior}
\end{figure}

Figure \ref{fig:prior} shows prior samples from the usual CPL prior used in most analyses, and the equivalent flexknot prior, and the corresponding posteriors from DES-5Y combined with DESI as kernel density estimates (KDEs).
The CPL prior includes the constraint that the value of $w$ at early times, $w_0+w_a$, is less than zero, to ensure a period of matter domination.
Clearly, the two priors are different, but the posteriors are so similar that it is difficult to see that there are two KDEs overlaid.
The different prior volumes will impact the evidence and tension values, in Appendix~\ref{apx:prior} it is shown that this effect is small.
The priors used in this work are listed in Table~\ref{tab:priors}.

\subsection{Nested bridge sampling}

The nested sampling approach used previously and in this work computes the Bayesian evidence for each combination of data and model.
However, in isolation, evidence is meaningless, and it is only with a pair that one can determine a Bayes factor, to which one may apply Jeffreys' scale \citep{jeffreys1939theory} to determine how to interpret the result, or combine with a model prior to determine the posterior odds.
An alternative method to compute the Bayes factor, besides two normal nested sampling runs, is to use nested bridge sampling.

Nested bridge sampling (NBS, \cite{chen2000monte, bridgetutorial}, Yallup et al. (in prep)) makes use of the inevitable similarity between the posteriors for the shared parameters between sampling runs with nested models.
This work consists entirely of nested models.
\lcdm{} is nested within $w$CDM with $w=0$, which are both nested within CPL, which are all nested within flexknot models.
Also, each cosmological model with $\dmb{}=0$ is nested within the same model with \dmb{} varying.
In theory, one could bridge sample from vanilla \lcdm{} to an $n=20$ flexknot with \dmb{}.
As a demonstration, let us outline nested bridge sampling to add \dmb{} only to an existing run.

Yallup (in prep) explains this method in detail and its application to toy and cosmological examples; let us recap the methodology here:
Suppose we wish to compute the Bayes factor between two models, $Z_2/Z_1$, where model 2 has an additional parameter, \dmb{}, compared to model 1.
The recipe is as follows: first, produce a set of posterior samples for model 1, with $\dmb{}=0$.
Then, sample the likelihood ratio, $\tilde{\mathcal L}$, using the posterior from the first run as part of an effective prior $\tilde\pi$:
\begin{equation}
    \tilde{\mathcal L}=\frac{\mathcal L(D|\theta, \dmb{})}{\mathcal L(D|\theta, \dmb{}=0)}\text, \quad 
    \tilde\pi = \pi(\dmb{}|\theta)\frac{\pi(\theta)\mathcal L(D|\theta, \dmb{}=0)}{Z_1(D)}\text,
    \label{eq:ratio}
\end{equation}
where $D$ is the data (DES-5Y alone or with DESI BAO), and $\theta$ are the appropriate other parameters for the cosmology in question.
The fraction in $\tilde\pi$ is simply the posterior density for model 1.
In practice, the posterior samples $\theta$ from the first run already encode the density $\tilde\pi$ up to the factor $\pi(\dmb{}|\theta)$.
Each sample is therefore augmented with a draw of \dmb{} from its prior, and the likelihood ratio $\tilde{\mathcal L}$ is evaluated at each augmented point.
A second sampling run then treats $\tilde{\mathcal L}$ as a likelihood and $\tilde\pi$ as a prior; its ``evidence'' is the Bayes factor $Z_2/Z_1$ directly.
These expressions arise from rearranging the Bayes factor:
\begin{equation} \label{eq:ratioratio}
    \begin{aligned}
        \frac{Z_2}{Z_1} &= \frac{\int\mathcal L(\theta, \dmb{})\pi(\theta, \dmb{})\,\mathrm d\theta\,\mathrm d(\dmb{})}{Z_1} \\
        &= \int\underbrace{\frac{\mathcal L(\theta, \dmb{})}{\mathcal L(\theta)}}_{\tilde{\mathcal L}}
        \underbrace{\pi(\dmb{}|\theta)\frac{\pi(\theta)\mathcal L(\theta)}{Z_1}}_{\tilde\pi} \,\mathrm d\theta\,\mathrm d(\dmb{}) \text,
    \end{aligned}
\end{equation}
where the data $D$ is suppressed from the likelihoods and evidences for readability, and the probability product rule has been used to factorise the prior, $\pi(\theta, \dmb{}) = \pi(\dmb{}|\theta)\pi(\theta)$.
\dmb{} has also been suppressed where it is fixed to zero.

The word ``nested'' should not be confused with the same in nested sampling, which refers to the onion of likelihood contours.
However, nested sampling may indeed be used to perform nested bridge sampling, but one may substitute Sequential Monte Carlo \citep{smc}, which is what is used here, for both the initial sampling run and bridging, abbreviated as SMC-NBS.
A possible disadvantage of SMC compared to nested sampling is that it does not report an error bar, the error bars here are from ten repeats with different initial random seeds.

\subsection{\textsc{JAX} reimplementation}

In addition to the pipeline from previous work, the likelihoods have also been ported to \textsc{JAX}.
This allows them to be sampled using \blackjax{} nested slice sampling, which makes use of GPU parallelisation \citep{yallup2025nested, metha, cabezas2024blackjax}.
In fact, the $n=2$ flexknot prior and posterior shown in Figure~\ref{fig:prior} is from the \textsc{PolyChord} pipeline, which uses \textsc{NumPy} and \textsc{Scipy} and 64-bit floating point precision, while the CPL result uses the \textsc{JAX} pipeline, with 32-bit floating point precision.
Reimplementation in \textsc{JAX} posed some additional challenges, for example, the adaptive \textsc{QUADPACK} method used to compute the integral over $\frac{H_0}{H(z)}$ is not available in \textsc{JAX.Scipy}, so trapezoidal integration was substituted.
Also, it was found that the precision setting for the Mahalanobis distance matrix multiplication had to be increased from its default level for consistency with the other pipeline.
Therefore, it is very reassuring that the results, which differ in hardware, sampler, 1-dimensional integration technique for proper motion distance, and floating point representation, are so similar.

One feature of \textsc{PolyChord} currently omitted from the \blackjax{} sampler is live point clustering.
Clustering is typically considered necessary for nested sampling with multi-modal posteriors, however, the \blackjax{} sampler has proven surprisingly effective on other problems.
It is the subject of ongoing research whether no clustering is genuinely a limitation, therefore, we restrict the application of the \textsc{JAX} pipeline to the unimodal \lcdm{} and CPL likelihoods only in this work.

\section{Results}\label{sec:results}

\begin{figure}
    \begin{center}
        \includegraphics[width=0.48\textwidth]{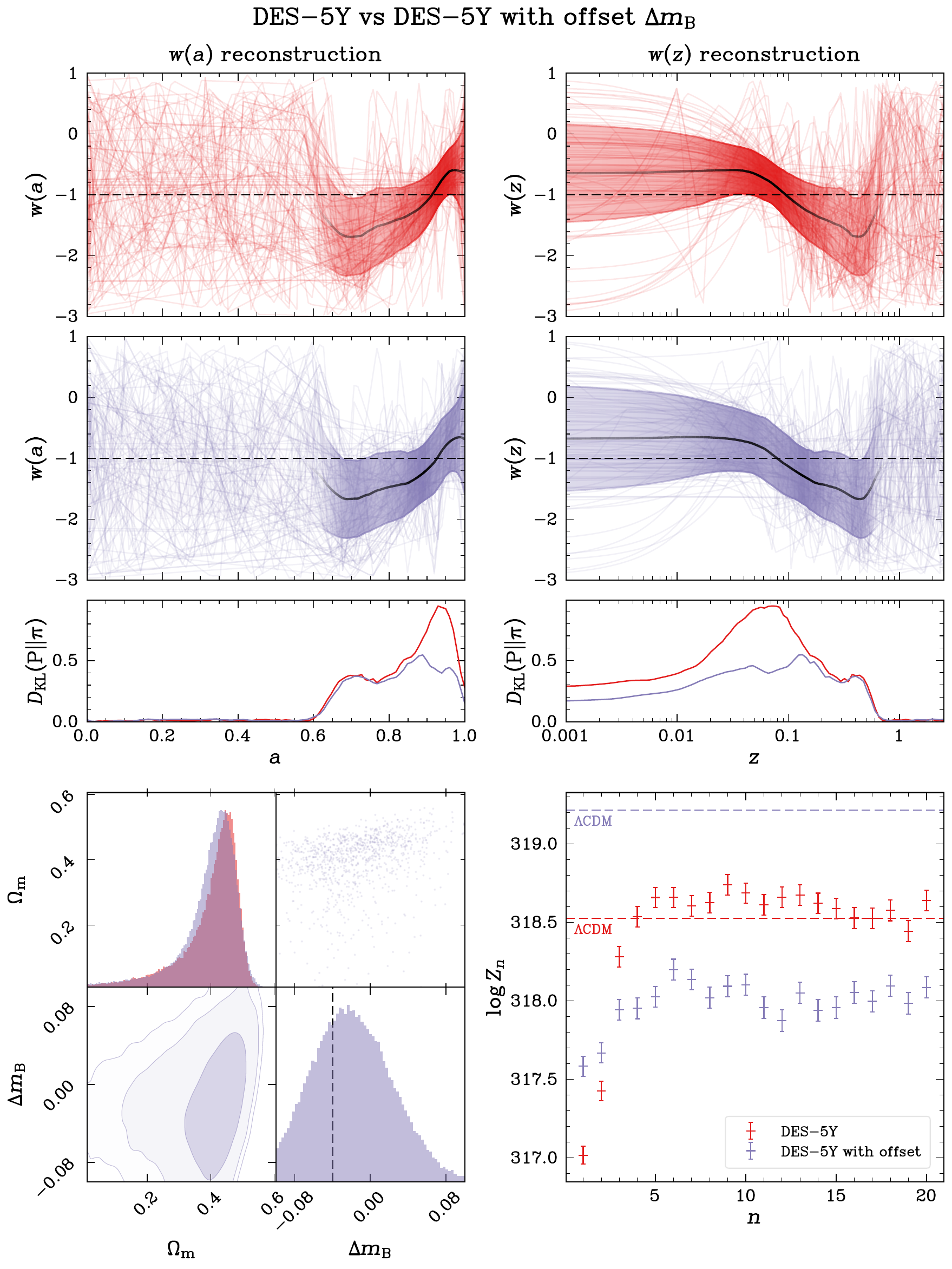}
    \end{center}
    \caption{
        Flexknot reconstruction of the dark energy equation-of-state parameter using DES-5Y supernovae only.
        In red is the standard likelihood, in lilac, the version with the \dmb{} offset for the low-redshift supernovae.
        The overall shape of the reconstructions are very similar, but the functional KL divergence and model evidences tell quite different stories.
        First, note that the high-$a$/low-redshift KL divergence lacks the peak just below $z=0.1$, which is to be expected as allowing those magnitudes to float up and down will naturally reduce their constraining power.
        Second, note that the evidence for \lcdm{} has increased with the offset, meanwhile, it has fallen for all flexknots with more than three knots.
        The Bayes factor between \lcdm{} and $w$CDM is similar between the two likelihoods, but $n=2$ is more disfavoured with the offset likelihood.
        Crucially, the evidence for \lcdm{} with the offset is greater than the flexknots without the offset.
        This suggests that the complexity demanded by the flexknot model is just as well, if not better, met by including this additional degree of freedom.\\
        Please note that the posteriors shown in the bottom-left panels do not include \lcdm{}, which is shown separately in Figure~\ref{fig:dmb}.
        With flexknots, \dmb{} is not well constrained.
    }\label{fig:des5y}
\end{figure}

\begin{figure}
    \begin{center}
        \includegraphics[width=0.48\textwidth]{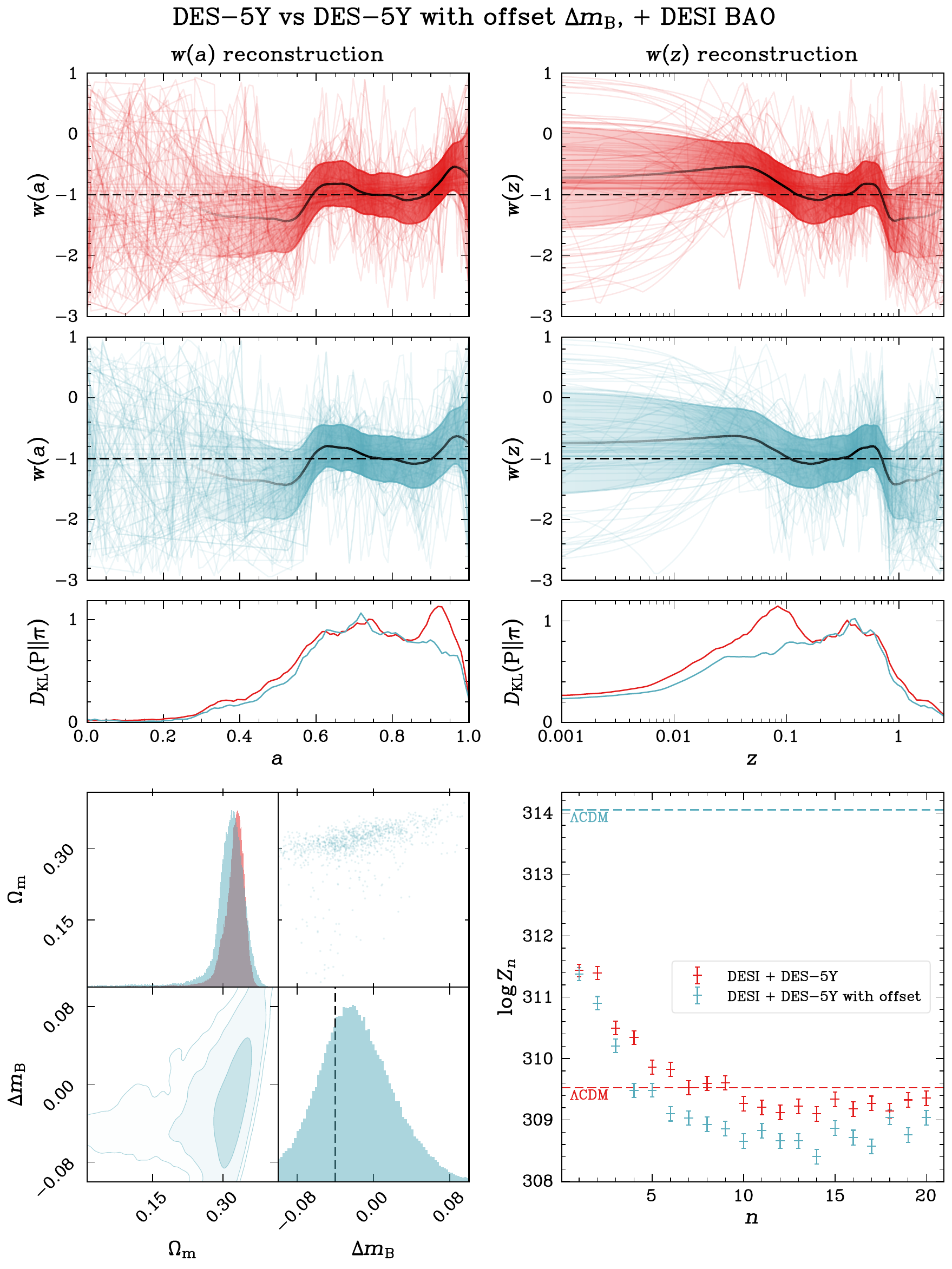}
    \end{center}
    \caption{
        Similar to Figure~\ref{fig:des5y}, with the addition of DESI BAO.
        This time, it is even clearer that \lcdm{} with the \dmb{} offset is the favoured model.
        Once again, the low-redshift KL divergence peak is lost with the additional parameter.
        Unlike the results with DES-5Y alone, the Bayes factor for \lcdm{} with \dmb{} over almost any other model is ``decisive''.
        Again, please note that the posteriors shown in the bottom-left panels do not include \lcdm{}, these are shown in Figure~\ref{fig:dmb}.
    }\label{fig:desidr2des5y}
\end{figure}

\begin{figure}
    \begin{center}
        \includegraphics[width=0.4\textwidth]{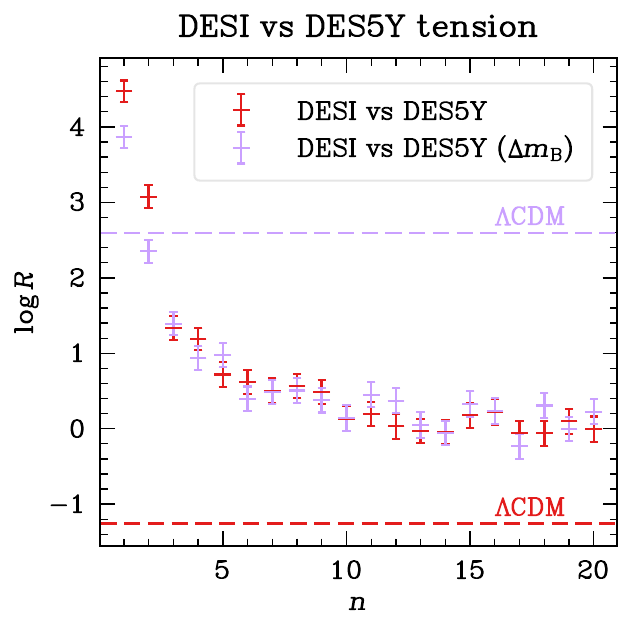}
    \end{center}
    \caption{
        Tension values between DES-5Y and DESI BAO, with and without the low-redshift offset.
        \lcdm{} is shown as horizontal dashed lines, for easy comparison with the other points.
        For \lcdm{} the tension has been reduced (more positive) significantly, while for $w$CDM and CPL, it has increased slightly.
        \lcdm{} with the offset is now on-par with CPL, though $w$CDM remains the model with the best dataset concordance.
        In contrast, without the offset, \lcdm{} is the most discrepant model of all.
        For three knots and above, the tension is similar with or without the offset.
    }\label{fig:tension}
\end{figure}

\begin{figure}
    \begin{center}
        \includegraphics[width=0.23\textwidth]{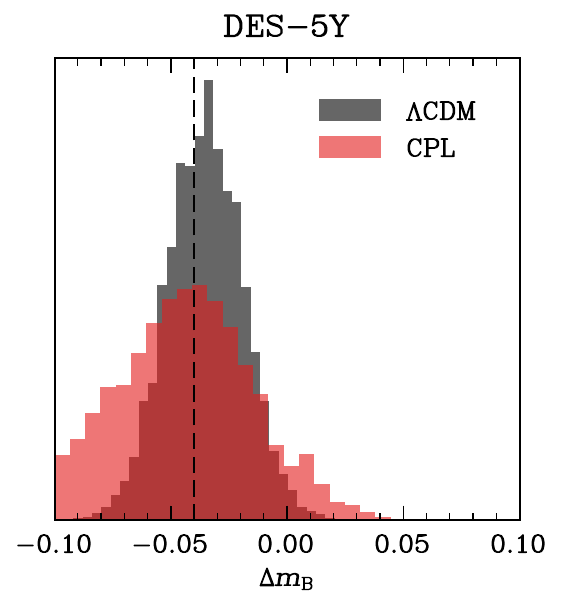}
        \includegraphics[width=0.23\textwidth]{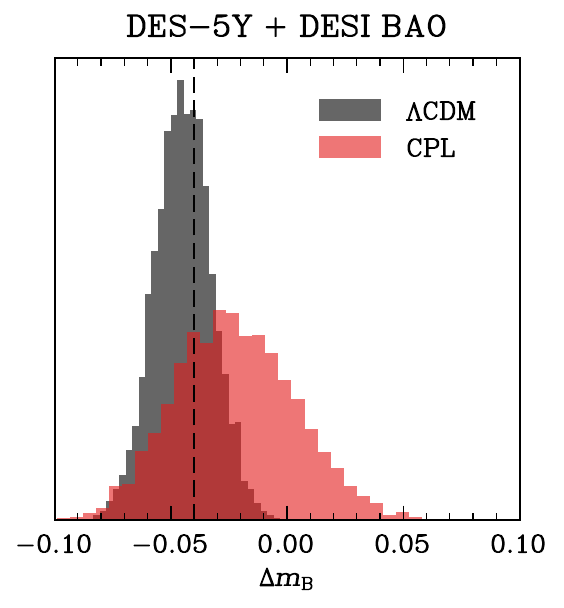}
    \end{center}
    \caption{
        Posterior histograms of \dmb{} for \lcdm{} and CPL.
        The left panel uses DES-5Y only, the right also includes DESI BAO.
        The prior is uniform over the domain of the plot.
        As predicted by \citet{georgedes5y}, its value is centred on $-0.04$.
        Note that the \lcdm{} posteriors (and the right CPL posterior) are well contained within the prior, therefore, the evidence which would have been found had a wider prior been used can be easily computed with the ratio of the prior volumes.
    }\label{fig:dmb}
\end{figure}

\begin{table}
    \caption{
        Values of \dmb{} for DES-5Y, along with the log-Bayes factor between the standard version ($\dmb{} = 0$) and with the offset.
        The Bayes factor is calculated in two ways, first by taking the ratio of the evidences from the nested sampling runs, and also by bridge sampling from $\dmb{}=0$ to the offset version.
        Bridge sampling is not attempted for the flexknot models, as more work is required to determine if this is viable.
        The nested sampling Bayes factors in the table are from the \textsc{JAX} pipeline; these are consistent with the \textsc{PolyChord} pipeline, which are not reported here.\\
        Both \lcdm{} values of \dmb{} are consistent with $-0.04$, and inconsistent with zero to over $2\sigma$.
        In contrast, both CPL and flexknot dark energy do not exclude zero, which is reflected in the Bayes factors.
        Of particular note is the large positive log-Bayes factor for \lcdm{} with DES-5Y + DESI BAO in favour of a low-redshift supernova offset.
        The Bayes factors from SMC-NBS are beyond error of those computed from two nested sampling runs, but not so much to affect any conclusions.\\
        The bottom section of the table updates the middle portion with DES-Dovekie.
        The log-Bayes factor between \dmb{} and vanilla \lcdm{} has almost halved, is similar to before with CPL, and consistent with zero for flexknots.
        Bridge sampling has not been attempted for the Dovekie data, as it has already been validated with DES-5Y.
        Curiously, the mean \dmb{} values are closer to $-0.04$ than before, particularly with flexknots.
    }\label{tab:des5y}
    \begin{center}
        \begin{tabular}[c]{|l|c|c|c|}
            \hline
            \multicolumn{4}{|c|}{DES-5Y}\\
            \hline
            & \dmb{} & $\log(\text{Bayes factor})$ & SMC-NBS \\
            \hline
            \lcdm{} & $-0.035 \pm 0.016$ & $0.755 \pm 0.098$ & $0.665 \pm 0.030$ \\
            CPL & $-0.042 \pm 0.028$ & $-0.067 \pm 0.127$ & $0.086 \pm 0.109$ \\
            flexknot & $-0.018 \pm 0.040$ & $-0.525 \pm 0.021$ & N/A \\
            \hline
            \multicolumn{4}{|c|}{DES-5Y + DESI BAO}\\
            \hline
            \lcdm{} & $-0.045 \pm 0.012$ & $4.140 \pm 0.182$ & $4.488 \pm 0.037$ \\
            CPL & $-0.022 \pm 0.026$ & $-0.859 \pm 0.217$ & $-0.738 \pm 0.018$ \\
            flexknot & $-0.017 \pm 0.038$ & $-0.396 \pm 0.055$ & N/A \\
            \hline
            \multicolumn{4}{|c|}{DES-Dovekie + DESI BAO}\\
            \hline
            \lcdm{} & $-0.038 \pm 0.013$ & $2.526 \pm 0.167$ & N/A \\
            CPL & $-0.033 \pm 0.024$ & $-0.457 \pm 0.200$ & N/A \\
            flexknot & $-0.035 \pm 0.027$ & $0.026 \pm 0.073$ & N/A \\
            \hline
        \end{tabular}
    \end{center}
\end{table}

\subsection{Flexknot reconstructions}
We begin with DES-5Y supernovae alone.
Figure~\ref{fig:des5y} shows the flexknot reconstruction of $w(a)$ both with and without the low-redshift offset.
From the KL divergence panels, it can be seen that the constraining power at low redshifts is reduced with the additional parameter; this is to be expected.
The model with the greatest evidence is \lcdm{} with the offset.
This suggests that the complexity picked up by the flexknot model, which brings models with four or more knots in line with \lcdm{}, is better explained by \dmb{}.
Only \lcdm{}, $w$CDM and CPL ($n=1$ and $n=2$ respectively) have greater evidence with the introduction of the offset; the reverse is true for three or more knots.
However, at least with DES-5Y alone, these suggestions must be caveated by the relatively small log-Bayes factors between evidences with the same $n$, which are at most around $0.7$.
In fact, Jeffreys' scale suggests that these differences are ``barely worth mentioning''.
We will return to this in Section~\ref{sec:priorwidth}.

When the DESI data are included, the combined reconstruction is shown in Figure~\ref{fig:desidr2des5y}.
With the BAOs, \lcdm{} with \dmb{} is still the favoured model, this time, with a log-Bayes factor of around four over almost all other models, crucially including standard \lcdm{}.
The offset has also removed the preference for low numbers of flexknots.
This strongly suggests that the original preference for dynamical dark energy can be better accounted for by this offset than it can by $w$CDM, CPL, or even flexknots.

The \textsc{JAX} pipeline produced \lcdm{} and CPL evidences in perfect agreement with those from \textsc{PolyChord}; in fact, it is those reported in Table~\ref{tab:des5y}.
These nested sampling runs completed in less than a second on a Google Compute Engine Nvidia L4 GPU, plus a few seconds of compile time.
This compares very well to tens of minutes on a 76-core CSD3 CPU, plus submission delay, and has the potential to transform the typical Bayesian's workflow.

It is important to assess how \dmb{} may affect the possibility of tension between DES-5Y and DESI BAO.
Once again, we follow \cite{paper1}, using the techniques developed in \cite{lemos, hergt, balancingact} via the $\log R$ statistic.
The results are shown in Figure~\ref{fig:tension}.
Without the offset, \lcdm{} is the most discrepant model, with the only negative $\log R$.
With the offset, \lcdm{} is now on par with CPL, while $w$CDM remains the model with the best concordance between the two datasets.
For three knots and above, the tension is similar with or without the offset.
This is reassuring, as it suggests that the offset is better able to explain the discrepancy between the datasets than flexknot dark energy.

\subsection{Effect of \dmb{} prior width}\label{sec:priorwidth}

From Figure~\ref{fig:dmb}, it can be seen that, for \lcdm{}, the posterior for \dmb{} is reasonably well contained within our chosen prior.
This allows us to examine the Bayes factors in more detail.

The uniform prior contains a factor of the reciprocal of the prior volume $V$, which, in the well contained case, persists into the Bayesian evidence.
This means that the evidence that \textit{would} have been found for a different, wider, uniform prior can be calculated as:
\begin{equation}
    \log Z \rightarrow \log\left(Z\times\frac{V_\text{narrow}}{V_\text{wide}}\right) = \log Z - \log\frac{V_\text{wide}}{V_\text{narrow}}\text.
    \label{eq:}
\end{equation}
Our prior of $\dmb\in[-0.1, 0.1]$ was chosen to be relatively narrow for two reasons: firstly, the nested sampling runtime is proportional to the KL divergence from prior to posterior; second, if the low-redshift systematic offset does exist and had a value beyond this range, it is unlikely that it would fall to third parties to investigate it, and frankly this should impact on our model prior.
For example, let us examine our conclusions if a prior ten times wider, $[-1.0, 1.0]$, had been chosen.
This would reduce log-Bayes factors by approximately $\log 10 = 2.30$.
Consider DES-5Y alone: the only positive log-Bayes factor is \lcdm{} at around $0.76$, which would be reversed with this more liberal prior.
In fact, we may reverse-engineer this argument to note that, had a prior $e^{0.76}\approx2.14$ times larger been chosen, the log-Bayes factor would have been precisely zero.
One may rightly criticise that post-hoc revisions to the prior are unwise, however, if this were all the data we had, it seems this would be enough to suggest that there is no practical evidence for the low-redshift offset, since it is so vulnerable to somewhat reasonable alternative priors.

However, now reintroduce the DESI BAO, with a log-Bayes factor of $\log Z = 4.140\pm0.127$ in favour of \dmb{}.
This time, one would have needed a prior at least sixty-two times wider to reverse the conclusion --- an offset of such magnitude is completely unreasonable.
Therefore, we will conclude that there is only evidence for the low-redshift supernova offset in DES-5Y if, and only if, it is combined with baryon acoustic oscillations, and \lcdm{} is correct.

\begin{figure}
    \begin{center}
        \includegraphics[width=\columnwidth]{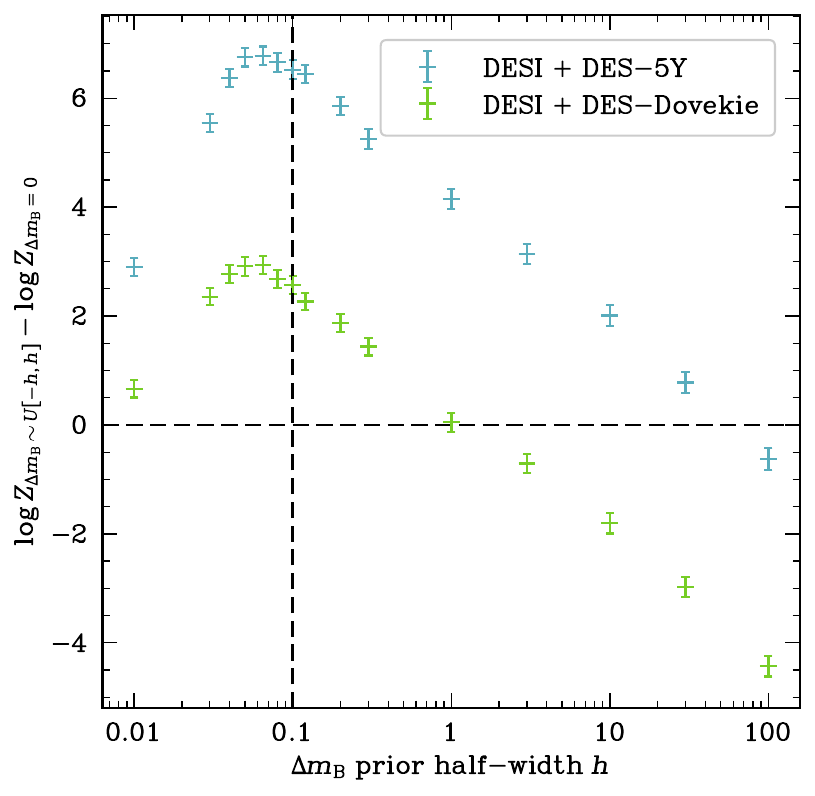}
    \end{center}
    \caption{
        Log-Bayes factor for \lcdm{} with \dmb{} over vanilla \lcdm{} as a function of the prior half-width $h$, where $\dmb{}\sim U[-h, h]$.
        The dashed vertical line indicates the prior used in this work, $h=0.1$.
        For both DES-5Y and DES-Dovekie combined with DESI BAO, the Bayes factor peaks very close to the chosen prior of $h=0.1$; narrower priors cut into the likelihood support, while wider priors are Occam-penalised.
        For $h > 0.1$, the decline is remarkably linear in $\log h$, consistent with the analytic $\log(V_\text{wide}/V_\text{narrow})$ scaling of Equation~\ref{eq:}, validating the back-of-the-envelope estimates of factors of sixty-two (DES-5Y) and thirteen (DES-Dovekie, Section~\ref{sec:dovekie}).
    }\label{fig:width}
\end{figure}

This argument is extended in Figure~\ref{fig:width}, which shows the log-Bayes factor as a function of the prior half-width for both DES-5Y and DES-Dovekie combined with DESI BAO.
The peak is very close to the chosen prior of $h=0.1$, and for wider priors the decline is strikingly linear in $\log h$, exactly as predicted by Equation~\ref{eq:}.
This confirms that the simple rescaling argument used to derive the factor of sixty-two is a good approximation.
This means that our prior is close to maximally informative: wide enough to encompass the likelihood, but narrow enough to avoid unnecessary Occam penalty.
However, the logarithmic decline is slow enough that the prior must be widened by over an order of magnitude before the conclusion is reversed.

\subsection{Nested bridge sampling Bayes factors}

Table~\ref{tab:des5y} also includes the SMC-NBS Bayes factors.
Error bars were estimated by running SMC-NBS ten times with different random seeds, while those from nested sampling were computed using \textsc{anesthetic}, which samples possible nested sampling volume compression histories \citep{anesthetic}.
It is interesting that all four of the NBS error bars are tighter than those from nested sampling.
The different approaches, in some cases, lie outside the error of each other, though not so significantly as to affect any conclusions.
For example, it is not surprising that the very small log-Bayes factors with CPL DES-5Y have opposite signs.
Crucially, the Bayes factor of the most interest, which happens to be the largest, of \dmb{} or no \dmb{} with DESI in \lcdm{}, is similarly large with NBS.
The SMC implementation also uses \blackjax{}, and similarly takes mere seconds to run.

\section{Update in light of DES-Dovekie}\label{sec:dovekie}

\begin{figure}
    \centering
    \includegraphics[width=0.48\textwidth]{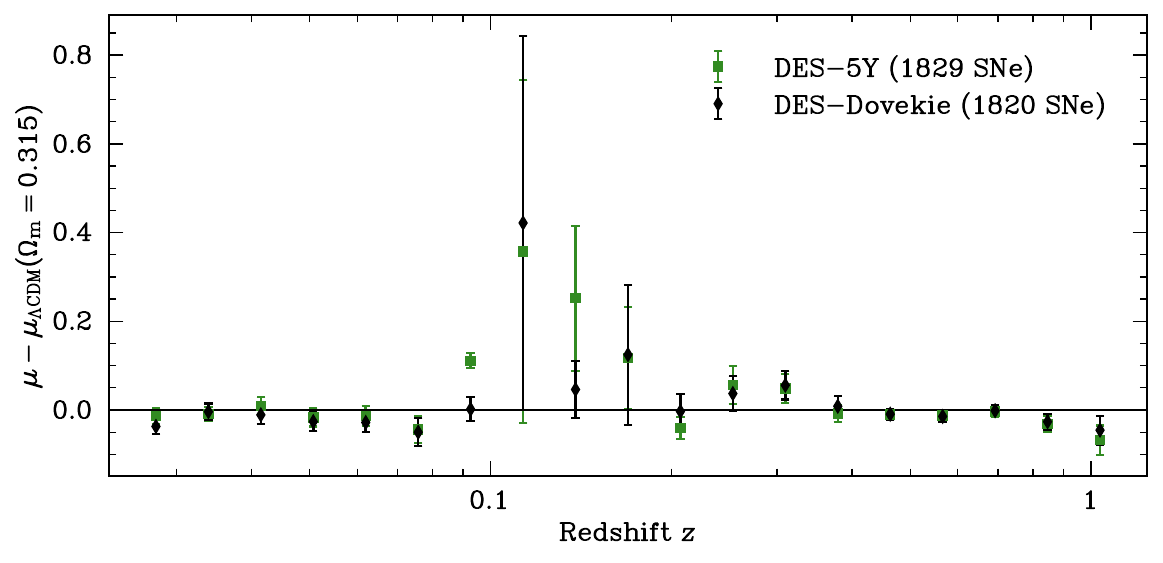}
    \caption{
        Hubble diagrams for DES-5Y and DES-Dovekie supernovae.
        To make the differences between the data more apparent, the data have been grouped into log-spaced redshift bins, and subtracted from the distance modulus from \lcdm{} with $\Omega_\mathrm m=0.315$.
        It is not immediately apparent that the low-redshift data have changed very much, in fact it is the intermediate redshifts that are most noticeably changed.
        Certainly, it is only by passing the new data through the Bayesian pipeline that the conclusions can be accurately updated.
    }\label{fig:dubble}
\end{figure}

Since submission at the end of September 2025, the DES collaboration released their DES-Dovekie reanalysis of the 5-year sample in November 2025 \citep{dovekie}, with improved photometric cross-calibration, additional calibration from recent white dwarf observations, and some bugfixes.
The new data are shown (binned) in Figure~\ref{fig:dubble}, along with DES-5Y for comparison.
It is not immediately apparent that the low-$z$ data have changed significantly, in fact, it is the intermediate redshifts which appear to have changed the most.
\citet{georgedes5y} and the previous sections of this work focussed on the low-$z$ supernovae, allowing the possibility of their being offset by \dmb{}.
However, all of the supernova likelihoods are marginalised over the absolute magnitude $M_\mathrm B$, which means that identical posterior plots, evidences, tensions and conclusions would be drawn by offsetting the complementary set of high-$z$ supernovae with the opposite sign.
It is not possible to look at the data as presented in Figure~\ref{fig:dubble} and accurately speculate how this will affect the evidence of \lcdm{} relative to dynamical models of dark energy.
Thus, the new data provided an opportunity to update the conclusions of this paper, and to revisit the performance of the \blackjax{} pipeline.

\begin{figure}
    \begin{center}
        \hordoverlay[0.98]{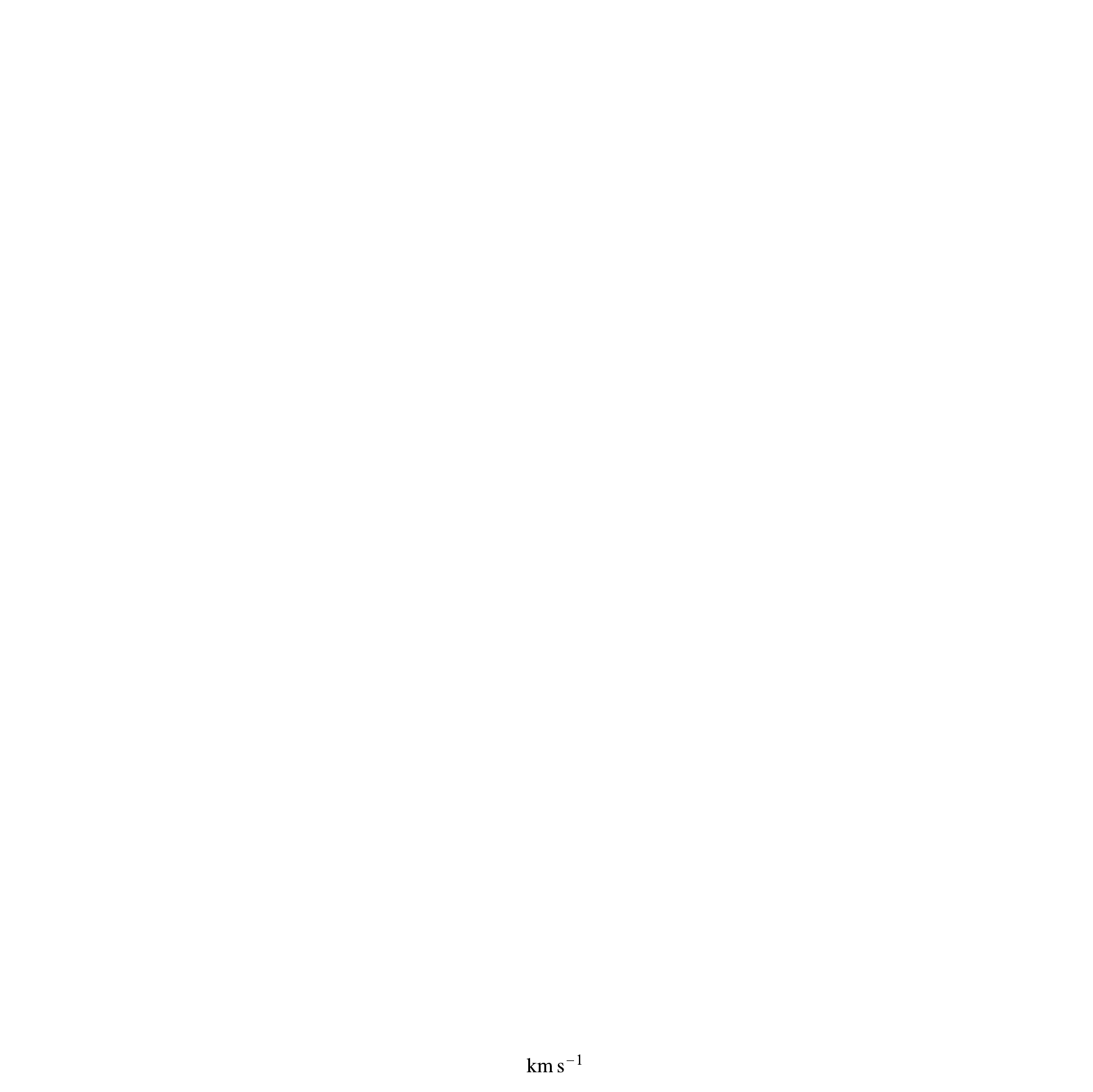}{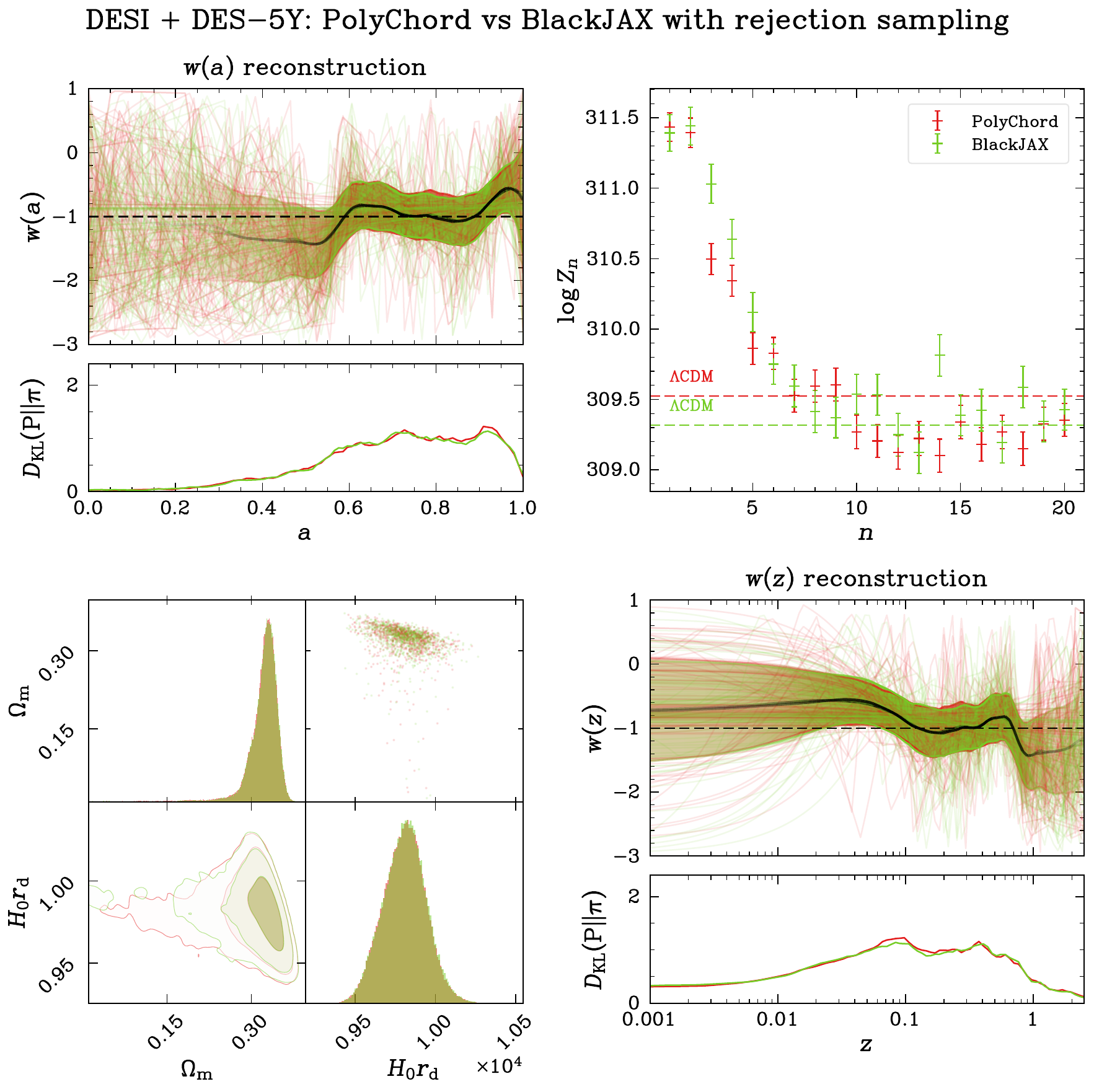}
    \end{center}
    \caption{
        Comparison of $w(a)$ reconstruction using DESI DR2 and DES-5Y using \pc{} and \blackjax{} nested slice sampling; \blackjax{} is using rejection sampling to enforce the sorting of the $a_i$ parameters, with up to twenty knots.
        The \blackjax{} results are statistically indistinguishable from those produced with \pc{}.
        Now that \blackjax{} is working correctly, the dominant cost of updating the results of this paper is no longer the sampling runtime, but detailed changes to the likelihood to account for the slightly different formatting of the Dovekie data.
    }\label{fig:rejection}
\end{figure}

\begin{figure}
    \begin{center}
        \includegraphics[width=0.23\textwidth]{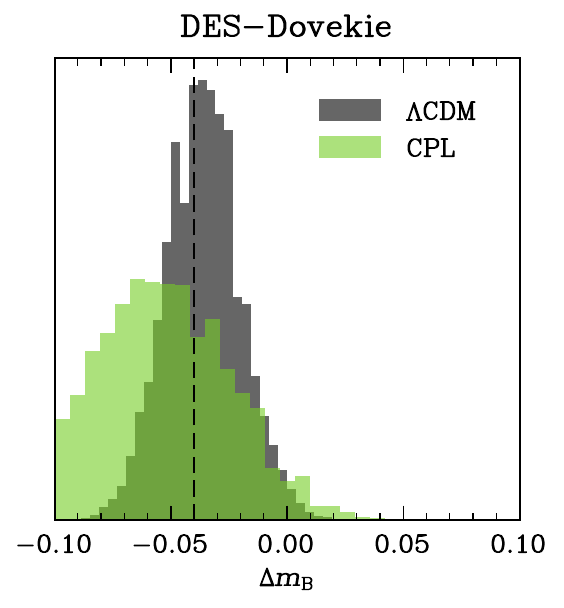}
        \includegraphics[width=0.23\textwidth]{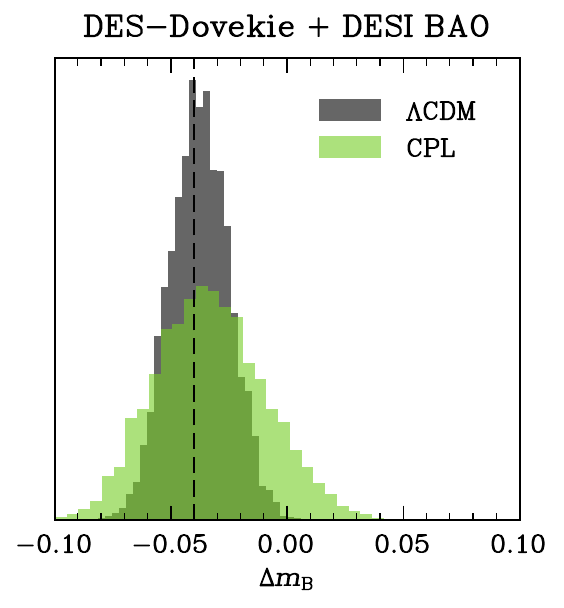}
    \end{center}
    \caption{
        Posterior histograms of \dmb{} for \lcdm{} and CPL.
        This updates Figure~\ref{fig:dmb} with DES-Dovekie.
        The left panel uses DES-Dovekie only, the right also includes DESI BAO.
        The prior is uniform over the domain of the plot.
        The distribution is still centred well away from zero, at around $-0.04$.
        }\label{c7fig:dmb}
\end{figure}

\blackjax{} has received updates in the meantime, and more thought has been given to the best way to handle the sampling of sorted parameters (the scale factors $a_i$ of the flexknots).
Previous attempts used reflection to address this challenge, if a slice sampling step extends beyond the region of parameter space where the $a_i$ are sorted, then they are swapped to fix the ordering, along with their corresponding $w_i$.
However, this does not take into account that the particles' covariance matrix is used by \blackjax{} to propose slice sampling directions, and this requires correction in order to be appropriate when it is possible for the parameters to be, essentially, swapped over.

There are two obvious alternatives: the more complex is to use a bijector to transform between a set of \textit{unsorted} values, which are sampled by \blackjax{}, and the actual sorted $a_i$, which are used by the likelihood.
Alternatively and more simply: one can instead initialise \blackjax{} with sorted samples, and define the prior density to zero for unsorted $a_i$; rejection sampling.
Both methods have proven successful in that they produce flexknot reconstructions identical to those using \pc{} up to at least twenty knots, with a tiny fraction of the runtime.
The DESI DR2+DES-5Y reconstruction using rejection sampling is compared to the equivalent \pc{} reconstruction in Figure~\ref{fig:rejection}; they are statistically identical (save for a spurious realisation of the evidence for three knots).
For comparison, the twenty-knot DESI DR2+DES-5Y nested sampling run took 5h24m52s on 76 icelake cores with the \pc{} pipeline.
With the same twenty knots running on an A100 GPU, \blackjax{} takes 1m43s including compilation time, 1m11s of which is actually spent sampling; the times are consistent to within a few seconds between the bijector and rejection sorting approaches, if anything, rejection sampling is a hair faster.
This means that the dominant cost of reanalysis with the DES-Dovekie update is no longer the sampling runtime; in fact, the detail changes required for the slightly different formatting of DES-Dovekie took longer to implement!

\begin{figure}
    \begin{center}
        \includegraphics[width=0.48\textwidth]{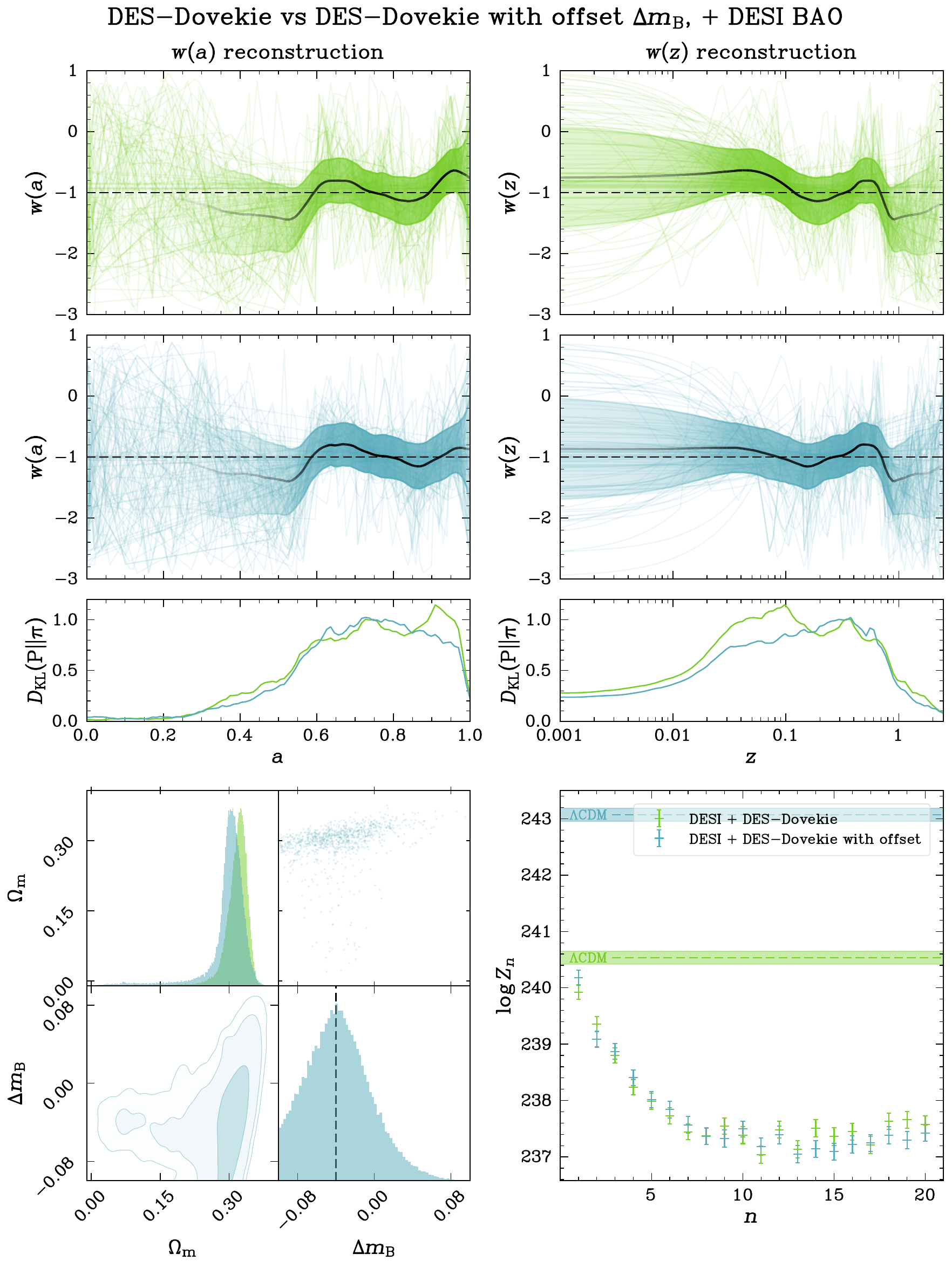}
    \end{center}
    \caption{
        Similar to Figure~\ref{fig:desidr2des5y}, using DES-Dovekie instead of DES-5Y, and \blackjax{} (with improvements to sorted parameter sampling) instead of \pc{}.
        A shaded region to indicate the error bar of the \lcdm{} evidence has also been added.
        Additionally, the error bars of \lcdm{} are now shown using shading.
        Aside from an overall rescaling of the evidences, the most notable difference is that the evidence for \lcdm{} without the additional offset has shifted upwards relative to all other evidences, which are otherwise very similar to how they were with DES-5Y.
        While \lcdm{} with the \dmb{} offset still has the greatest evidence of all models, vanilla \lcdm{} now has greater evidence than any number of flexknots.
        Interestingly, the \dmb{} posterior peaks at almost exactly $-0.04$!
    }\label{fig:jovekie}
\end{figure}

\begin{figure}
    \begin{center}
        \includegraphics[width=\columnwidth]{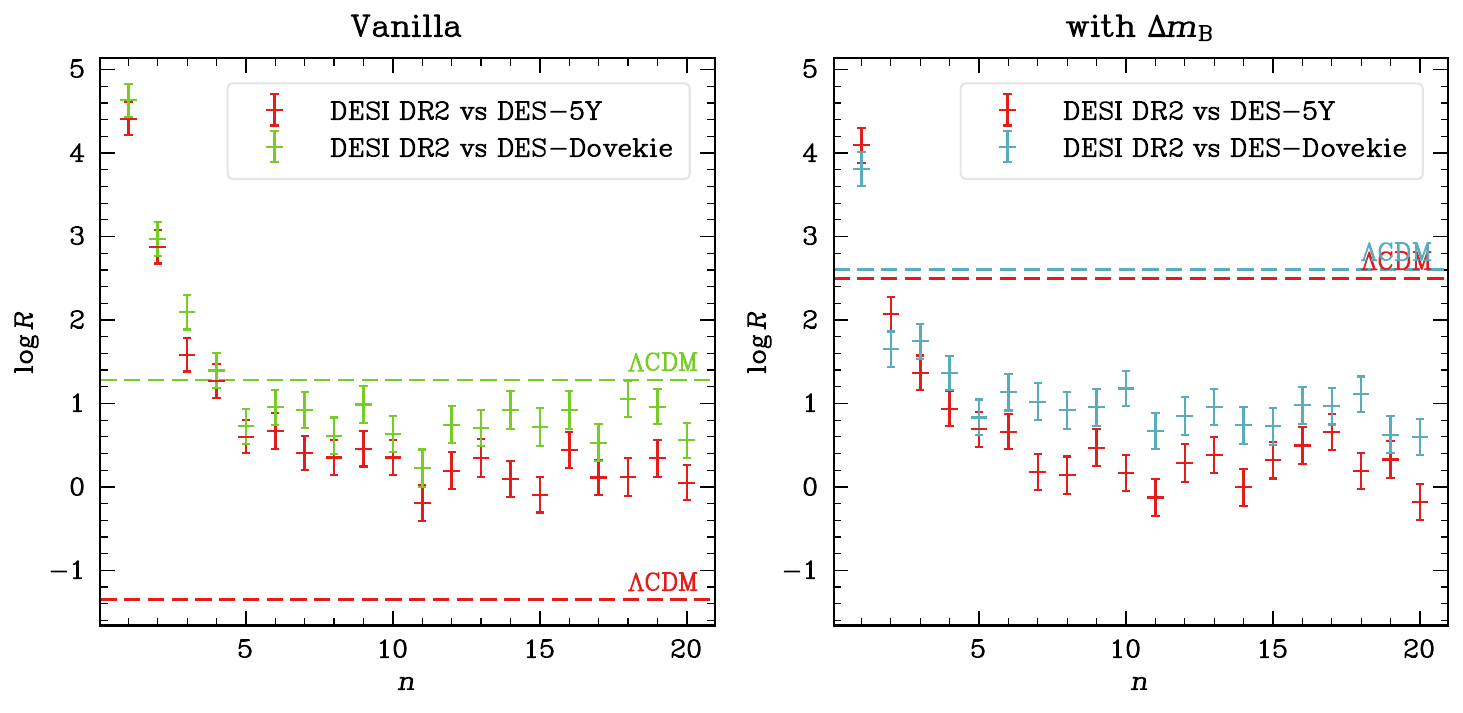}
    \end{center}
    \caption{
        Comparison of tension statistics with DES-5Y or DES-Dovekie versus DESI DR2.
        Vanilla refers to the unmodified supernova datasets, as opposed to including the low-redshift \dmb{} offset.
        The red data in this figure are the red and mauve data from Figure~\ref{fig:tension}, the updated data are blue and green, to match Figure~\ref{fig:jovekie}.
        It can be seen that tension is reduced (more positive $\log R$) with vanilla \lcdm{} with DES-Dovekie, though one to three knots still reduces the tension further.
        Tensions with flexknots are very similar, though perhaps overall slightly reduced, across both data releases.
        The addition of \dmb{} still further reduces tension with \lcdm{}, but to a far lesser extent with DES-Dovekie than it did with DES-5Y.
        Indeed, the tension is still less for DES-5Y with \dmb{}, versus DES-Dovekie with the vanilla model.
    }\label{fig:jovekietension}
\end{figure}

Convinced that \blackjax{} is now up to the task of producing posteriors and evidences consistent with \pc{}, let us return to the DES-Dovekie data.
Figures~\ref{c7fig:dmb} and \ref{fig:jovekie} update Figures~\ref{fig:dmb} and \ref{fig:desidr2des5y} respectively with the new supernova data, and the updated tension $\log R$ statistics are shown in Figure~\ref{fig:jovekietension}.
First, the posteriors of \dmb{}, both for \lcdm{} (Figure~\ref{c7fig:dmb}) and flexknots (panel within Figure~\ref{fig:jovekie}), remain centred close to $-0.04$, arguably even closer than before for the flexknots, and still significantly non-zero.
The updated values of \dmb{} and Bayes factors are reported in Table~\ref{tab:des5y}.
The log-Bayes factor between \dmb{} and vanilla \lcdm{} has almost halved, is similar to before with CPL, and consistent with zero for flexknots.
This suggests that DES-Dovekie has gone some way to reducing the evidence for an offset in the supernova data.
Following the same argument as Section~\ref{sec:priorwidth}, the prior on \dmb{} would need to be at least thirteen times wider to reverse the conclusion.
This is consistent with what we see in Figure~\ref{fig:width}.
While this is more reasonable than the sixty-two times wider with DES-5Y, one would still be allowing for the possibility of a systematic offset \dmb{} of $\mathcal O(1)$; thus, we still conclude that the data favour \dmb{} over vanilla \lcdm{}.
Curiously, the mean \dmb{} values are closer to $-0.04$ than before, particularly with flexknots.

The $w(a)$ reconstructions are of identical character to before.
Besides an overall rescaling, the evidences also follow a similar character to DES-5Y; the notable exception being vanilla (no \dmb{}) \lcdm{}, which has moved up above all flexknot models, though still below \lcdm{} with the \dmb{} offset.
This means that the Bayes factor for \lcdm{} with a systematic low-$z$ issue is less significant than it was with DES-5Y.

The tension statistics offer a similar narrative.
Tension is reduced ($\log R$ is more positive) with vanilla \lcdm{} for DES-Dovekie versus DESI DR2 compared to how it was with DES-5Y; tension values for all other flexknot models are statistically unchanged.
This means that the flexknot models, besides those with 1--3 knots, no longer offer a tension reduction.
The tension is still further reduced with the addition of \dmb{}, but not as significantly as before.
Comparing across the two panels of Figure~\ref{fig:jovekietension}, it can be seen that \lcdm{} DES-Dovekie does not quite increase $\log R$ to the level it was at for DES-5Y with \dmb{}.

In summary, DES-Dovekie partially alleviates the tension between the low- and high-$z$ DES supernovae, but doesn't eliminate it.
The evidence for a systematic offset is weaker, but still present when combined with DESI BAO.
I leave the determination of the root cause of this tension to the capable hands of future work.

\section{Conclusions}\label{sec:conclusions}

In this work, we found that there was substantial Bayesian evidence for a low-redshift supernova systematic in DES-5Y, when it was combined with DESI BAO.
However, without the BAO, the claim was unsubstantiated, though the posterior on the offset value indeed agreed with the $-0.04$ of \citet{georgedes5y} and excluded zero to $2\sigma$.
Crucially, the Bayesian evidence favoured offset \lcdm{} over flexknot dark energy with or even without the offset, so we concluded that the systematic was a better model than dynamical dark energy.
We accept that this approach is limited, and adjusting a subset of apparent magnitudes post bias correction does not constitute a sensible supernova catalogue.
This is because bias corrections are derived from simulations assuming the original apparent magnitudes; post-hoc adjustments invalidate these corrections, which should ideally be recalculated from scratch.
Nevertheless, the options were clear: either there was a systematic issue with the DES-5Y supernovae, or dark energy really was dynamical.

It was also found that the \lcdm{} tension between DES-5Y and DESI BAO was significantly reduced with \dmb{}, while $w$CDM and CPL tensions were slightly increased.
$w$CDM remained the model with the best agreement.

The release of DES-Dovekie alters the picture.
Combining with DESI BAO, vanilla \lcdm{} now has greater evidence than all flexknot models, so evidence for dynamical dark energy is weakened.
\lcdm{} with the offset remains the model with the greatest evidence overall, and this conclusion is robust to reasonable changes in the prior on \dmb{}.
Dataset tension is also reduced for vanilla \lcdm{}, though both the offset and models with few knots can still reduce tension further, and $w$CDM is still the model with the best agreement both with and without \dmb{}.

The alternative \textsc{JAX} pipeline originally produced \lcdm{} and CPL posterior samples and evidences in excellent agreement with those from our original \textsc{PolyChord}-powered approach.
It is satisfying to have demonstrated that the \blackjax{} pipeline has now reached maturity, and that the DES-Dovekie reanalysis could be done for a fraction of the computational cost of the \pc{} runs, with no reduction in quality.
The reduced sampling time to mere seconds has the potential to be transformative for the Bayesian workflow.
We also find that Bayes factors obtained using nested bridge sampling are sufficiently similar to those obtained by the ratio of nested sampling evidences. Therefore, we hope that this approach can be trialled more widely.

\section*{Acknowledgements}

ANO and WJH were supported by the research environment and infrastructure of the Handley Lab at the University of Cambridge.
This work was performed using the Cambridge Service for Data Driven Discovery (CSD3), part of which is operated by the University of Cambridge Research Computing on behalf of the STFC DiRAC HPC Facility (\url{www.dirac.ac.uk}).
The DiRAC component of CSD3 was funded by BEIS capital funding via STFC capital grants ST/P002307/1 and ST/R002452/1 and STFC operations grant ST/R00689X/1.
DiRAC is part of the National e-Infrastructure.
WJH was supported by a Royal Society University Research Fellowship.
The authors thank Toby Lovick for useful correspondence regarding SMC-NBS.

\section*{Software and Data Availability}

The \textsc{python} pipeline in this work made use of \textsc{NumPy} \citep{numpy}, \textsc{SciPy} \citep{scipy}, and \textsc{pandas} \citep{pandaszenodo, pandaspaper}.
The GPU-accelerated pipeline was written in \textsc{JAX} \citep{jax2018github}, and supported by the Google Cloud research credits program, with the award GCP397499138.
The nested sampling chains were analysed using \textsc{anesthetic} \citep{anesthetic}; plots were produced in \textsc{matplotlib} \citep{matplotlib}, using the \textsc{smplotlib} template created by \citet{smplotlib}.
The \textsc{Python} and \textsc{JAX} pipelines and nested sampling chains used in this work can be obtained from Zenodo \citep{zenodo}.



\bibliographystyle{mnras}
\bibliography{desi3} 




\appendix

\section{Analytic marginalisation over \dmb{}}\label{apx:marginalisation}

This appendix contains an extension to the analytical marginalisation in \cite{paper1} and \cite{paper2}.
In those works, it was demonstrated that the constant offset from $M_\mathrm B$ could be analytically marginalised from the SNe likelihoods.
Since this value affects all terms in the data vector, there is essentially an accompanying mask of all ones.
In fact, the algebra is identical for any mask, assuming that the likelihood tends to zero at the limits of the prior, so \dmb{} could be marginalised over by replacing the covariance matrix:
\begin{equation}
    \Sigma^{-1} \rightarrow \Sigma^{-1} - \frac{\Sigma^{-1} \mathbfit s^T\mathbfit s\Sigma^{-1}}{\mathbfit s^T\Sigma^{-1}\mathbfit s} \text,
\end{equation}
and adjusting the normalisation by a factor of $\frac 1 {V_\mathbfit s}\sqrt{\frac{2\pi}{\mathbfit s^T\Sigma\mathbfit s}}$, where $V_\mathbfit s$ is the prior volume of the offset, \dmb{} in this case.
However, this introduces two challenges.
First, each one of these marginalisations introduces an additional zero eigenvalue to the inverse covariance matrix, reducing its rank by one.
This introduces precision issues, in particular when working in 32-bit floating point.
Secondly, the posterior of \dmb{} is of interest, and while it is possible to then sample it from the posterior of the other parameters, it is more straightforward to simply include it as a parameter during nested sampling, as this poses no challenge for our nested slice sampling tools.

\section{Effect of CPL prior on evidence and tension}\label{apx:prior}

This appendix investigates the impact of the difference between the flexknot and CPL priors on their Bayesian evidences and the tension ratio.
The Bayesian evidence may be split into two terms, the average log-likelihood over the posterior, and the KL divergence from prior to posterior:
\begin{equation}
    \log Z = \langle\log\mathcal L\rangle_\mathcal P - \mathcal D_\mathrm{KL}(\mathcal P||\pi)\text.
\end{equation}
In the uniform prior case, the KL divergence may be further simplified to:
\begin{equation}
    \begin{aligned}
        \mathcal D_\mathrm{KL}(\mathcal P||\pi) &= \int \mathcal P(\theta)\log\frac{\mathcal P(\theta)}{\pi(\theta)}\,\mathrm d\theta\\
        &= \int\mathcal P(\theta)\log\mathcal P(\theta)\,\mathrm d\theta - \int\mathcal P(\theta)\log\pi(\theta)\,\mathrm d\theta\\
        &= \int\mathcal P(\theta)\log\mathcal P(\theta)\,\mathrm d\theta + \log V_\pi\text,
    \end{aligned}
\end{equation}
where $V_\pi$ is the prior volume.
Assume that the two posteriors are very similar, as is the case for CPL and $n=2$ flexknots, then the posterior-averaged log-likelihoods will be approximately equal.
From Table~\ref{tab:priors}, the flexknot prior volume is 16, while the CPL prior volume is 15.5.
Therefore, the CPL log-evidences are expected to be $\log\frac{16}{15.5}\approx0.03$ greater than those for $n=2$ flexknots.

Consider the expression for the tension ratio:
\begin{equation}
    \log R = \log Z_\mathrm{SNe+BAO} - \log Z_\mathrm{SNe} - \log Z_\mathrm{BAO}\text.
\end{equation}
It can be seen that the tension ratio for CPL is expected to be approximately $0.03$ less than the equivalent flexknot.
This difference is not negligible, but small compared to the sampling uncertainty in the evidences themselves, see Table~\ref{tab:des5y} and Figure~\ref{fig:tension} for examples.


\bsp	
\label{lastpage}
\end{document}